# The Distributional Impact of Inflation in Pakistan: A Case Study of a New Price Focused Microsimulation Framework, PRICES

Cathal O'Donoghue*, Beenish Amjad***, Jules Linden* ** , Nora Lustig ***, Denisa M. Sologon**, Yang Wang***

* University of Galway; ** Luxembourg Institute for Social and Economic Research, *** CEQ Institute, Tulane University

**Abstract**

This paper develops a microsimulation model to simulate the distributional impact of price changes using Household Budget Survey data, income survey data and an Input Output Model. The primary purpose is to describe the model components. The secondary purpose is to demonstrate one component of the model by assessing the distributional and welfare impact of recent price changes in Pakistan. Over the period of November 2020 to November 2022, headline inflation 41.5%, with food and transportation prices increasing most. The analysis shows that despite large increases in energy prices, the importance of energy prices for the welfare losses due to inflation is limited because energy budget shares are small and inflation is relatively low. The overall distributional impact of recent price changes is mildly progressive, but household welfare is impacted significantly irrespective of households' position along the income distribution. The biggest driver of welfare losses at the bottom of the income distribution was food price inflation, while inflation in other goods and services was the biggest driver at the top. To compensate households for increased living costs, transfers would need to be on average 40% of pre-inflation expenditure, assuming constant incomes. Behavioural responses to price changes have a negligible impact on the overall welfare cost to households.

1. **Introduction**

With the resurgence of inflation, a growing interest in the use of price related environmental policy, and the interaction between these forces with indirect taxation, there is an increasing need to be able to evaluate the distributional impact of policy and economic changes. While there is a large and historic literature in these related fields, much of the work has been undertaken in a disjoint way. This paper describes the development of a framework to simulate the impact of price related policies, taking Pakistan as a case study.

The PRICES model (**P**rices, **R**evenue Recycling, **I**ndirect Taxation, **C**arbon, **E**xpenditure **S**imulation model) is able to simulate the impact of price increases from multiple sources, including those due to external inflationary shocks, and indirect and environmental taxes changes. It can also simulate policies to compensate households for increased living costs and estimates households' behavioural response to income and price changes, and the impacts on household welfare. A central contribution of PRICES is its ability to account for interactions between indirect and environmental taxes, inflation, compensation measures and household behaviour. In combining multiple modules and standardized data sources, the model provides a scalable and flexible framework to estimate the welfare impact of a wide range of price shocks, and is particularly apt for comparative research, across developed and developing countries. The purpose of this paper is to describe the current version of the PRICES model with all its components, and to showcase one component using the recent surge in inflation in Pakistan as a case study.

While microsimulation frameworks have been developed for cross-sectional income related policy, such as EUROMOD (Sutherland and Figari, 2013), and for inter-temporal analysis, such as the LIAM2 framework (De Menten et al., 2014) and the income generation model (Černiauskas et al., 2022), there is a gap in the sphere of price-related policy. Some studies incorporated price inflation to understand the impact of fiscal drag (Immervoll, 2005; Levy et al., 2010; Leventi et al., 2024) or the distributional and welfare impact of inflation (Sologon et al., 2022b). Others model the impact of indirect (Decoster et al., 2010; Brunetti and Calza, 2015; O'Donoghue et al., 2018) or environmental tax changes (Cornwell and Creedy, 1996; Grainger and Kolstad, 2010; Linden et al., 2023), sometimes including transfer-based revenue recycling (Feindt et al., 2021; Berry, 2019). Existing scalable microsimulation models generally focused on price changes due to indirect taxation (Akoğuz et al., 2020; Amores et al., 2023a), or environmental taxation and simple revenue recycling (Feindt et al., 2021; Steckel et al., 2021). These models however ignore the interplay between VAT, ad valorem, excise taxes and environmental taxes, do not estimate behavioural responses, or are limited to simple revenue recycling schemes. Information on incomes, direct tax and transfers is needed to simulate more sophisticated revenue recycling schemes. This information is typically not included in expenditure surveys. Statistical matching is often undertaken to link income and expenditure surveys to simulate expenditure and income related policies, (Decoster et al., 2011; Capéau et al., 2014). Recently, the EUROMOD team developed an indirect tax tool (ITT) that is able to assess distributional impacts of indirect taxation (Akoğuz et al., 2020).

The EUROMOD ITT however does not link household expenditure information to input output (IO) tables and includes limited behavioural responses. IO tables are required to model externality-correcting taxes, such as carbon taxes, as these taxes impact the cost of inputs, particularly energy, used in the production process of all goods and services. Carbon taxes are levied in relation to quantities of $CO_2$ emissions released during the production and consumption of goods and services. As expenditure datasets commonly do not include prices or quantities, this requires information on energy prices sourced from external sources or directly from IO databases.

As environmental taxes aim to reduce "social bads", models assessing their distributional impacts should allow for behavioural responses (Hynes and O'Donoghue, 2014). The aim of environmental taxation is to adjust prices to reflect negative externalities to steer production and consumption choices (Pigou, 1920; Stiglitz et al., 2017). In a modelling environment, this means that models need to estimate the level of pollution associated with the production and consumption of goods, to adjust the price of these goods accordingly, and estimate the extent to which households respond to these price changes. Estimates of household's responsiveness to price changes also form the basis of estimates of the welfare effects of policies, such as the Compensating Variation (CV). CV gives the monetary value required to maintain households' utility following a price increase. This is a particularly workable indicator for policymakers looking to compensate households.

Assessing the distributional impact of external and policy-driven price changes has increased policy relevance over the last years. The war in Ukraine drove energy price hikes, with broader price increases premediated by the COVID-19 crisis. Geopolitical volatility in the Middle East and South China Sea raise further concerns of price volatility in key inputs, such as energy and semi-conductors. Secondly, countries are implementing and raising carbon prices, including major economies such as the EU, Australia, Japan, California, and China. Distributional impacts of price changes are a key concern for political stability and social cohesion. Regarding the carbon pricing and indirect tax reform, distributional concerns are relevant for political acceptability of reforms, and regressive impacts can lead to major political opposition (Dechezleprêtre et al., 2022; Douenne and Fabre, 2022). Further, there is a push to reform energy taxes and subsidies to align energy prices with their carbon content, as exemplified by a recent proposal by the European Commission. This further exemplifies the need for integrated frameworks that jointly assess the distributional impact of multiple taxes.

Adequate policy responses to price changes resulting from economic and geopolitical volatility and a socially just transition towards low-carbon economies requires comprehensive tools that can assess the burden of prices for different households. This paper develops a data analytical tool to improve the design of policies that can both maintain living standards and deliver environmental goals whilst reducing the distributional impact.

The goal of this paper is to develop a novel and scalable analytical framework to assist policy makers in the design of better price-related policy. Our aim is to apply this framework in different continents at different stages of development and with different priorities in terms of mitigation measures. In this paper, we use the case study of recent inflation in Pakistan to illustrate the functioning of the behavioural and welfare component of the model. As the primary purpose of the paper is to introduce the PRICES model, the other functions of the model are described for completeness.

This paper describes in detail the development of a model that can jointly simulate distributional impacts of inflation, indirect taxation, carbon taxation, revenue recycling, and household behavioural responses. We describe the methodological context of environmental taxation microsimulation in section 2, and methodological issues in terms of modelling pollution and data in section 3. The distributional impact of inflation is simulated in section 4. Section 5 concludes.

2.  **Literature review**

There is a relatively extensive literature on modelling the distributive impact of indirect and environmental taxes.

Capéau et al. (2014) and O'Donoghue (2021) review the use of microsimulation for the simulation of price related issues. Most of the literature focuses on the distributional impact of indirect taxation (Decoster et al., 2010; O'Donoghue et al., 2018; Harris et al., 2018; Maitino et al., 2017; Symons, 1991).

A substantial literature uses microsimulation models for indirect taxation analysis in OECD countries including Australia (Creedy, 2001, Chai et al., 2021), Belgium (Decoster and Van Camp, 2001), Italy (Liberati, 2001; Brunetti and Calza, 2015; Gastaldi et al., 2017; Curci et al., 2022), Ireland (Madden, 1995; Leahy et al., 2011; Loughrey and O'Donoghue, 2012; Lydon, 2022), Greece (Tsakloglou and Mitrakos 1998; Kaplanoglou and Newbery, 2003), USA (Toder at al., 2013; Jaravel, 2021) and Germany (Kaiser and Spahn, 1989; Watt, 2022).

Indirect taxation, because they are collected at the point of sale and do not require elaborate administrative systems, are easier to collect than income taxes and therefore often forms a higher share of tax revenues in developing countries. Atkinson and Bourguignon (1991) found that much of the redistribution in the existing Brazilian system in the 1980s relied on instruments that were less important in OECD countries, where indirect taxes, subsidies and the provision of targeted non-cash benefits (such as public education and subsidised school meals) were found to be more important. Given the important share of tax revenue provided by indirect taxes and the availability of household budget survey data, the microsimulation modelling of indirect taxation is, and has been for a long time, a focus of developing and transition countries (Harris et al., 2018) such as Pakistan (Ahmad and Stern, 1991), Hungary (Newbery, 1995), Romania (Cuceu, 2016), Serbia (Arsić and Altiparmakov, 2013), Uruguay (Amarante et al., 2011), Guatemala, (Castañón-Herrera and Romero, 2011) and Chile (Larrañaga et al., 2012).

While most papers have focused on single country analyses, there is an increasing literature looking at indirect taxes in a comparative context (O'Donoghue et al., 2004; Decoster et al., 2010, 2011; Amores et al., 2023a).

Many of these papers focus on indirect taxes only, given that income data in household budget surveys is not always of sufficient quality to model direct taxes. In some cases as in the UK, it is possible to simulate both direct and indirect taxation (Redmond et al., 1998), but more often than not, there is a need to statistically match data from a budget survey into an income survey in order to model both direct and indirect taxes (Maitino et al., 2017; Akoğuz et al., 2020). Picos-Sánchez and Thomas (2015) undertook comparative research looking at joint direct and indirect tax reform in a comparative context.

Indirect taxes explicitly targeting environmental pollution, such as carbon taxation, have developed in parallel (Hynes and O'Donoghue, 2014; Cornwell and Creedy, 1996; Symons et al., 1994) and occasionally in joint analyses with indirect taxes (Decoster, 1995; Amores et al., 2023a).

Hynes and O'Donoghue (2014) provide a review of the wider literature of the use of microsimulation models for environmental policy. The distributional implications of carbon taxes have been analysed by O'Donoghue (1997) and Callan et al. (2009) in Ireland, Hamilton and Cameron (1994) in Canada, Labandeira and Labeaga, (1999), Labandeira et al. (2009) and García-Muros et al., (2017) in Spain, Bureau (2011), Bourgeois et al. (2021), Giraudet et al. (2021) and Berry (2019) in France, Casler and Rafiqui (1993), Grainger and Kolstad (2010), and Mathur and Morris (2014) in the USA, Symons (1994) and Symons et al. (2002) in the UK, Yusuf and Resosudarmo (2015) in Indonesia, Kerkhof et al. (2008) and Kerkhof et al. (2009) in the Netherlands, Bach et al. (2002) and Bork (2006) in Germany, Mardones and Mena (2020) in Chile, Chen (2022), Jiang and Shao (2014) and Zhang et al. (2019) in China, Poltimäe and Võrk (2009) in Estonia, Cornwell and Creedy (1996) in Australia, Rosas-Flores et al., (2017) and Renner et al. (2018) in Mexico, and Vandyck and Van Regemorter (2014) in Belgium. Multiple studies estimate household behavioural responses to carbon pricing (Cornwell and Creedy, 1996; Labandeira and Labeaga, 1999; Renner et al., 2018; Tovar Reaños and Lynch, 2023). Only few studies include the behavioural response to revenue recycling and rebound effects (Ravigné et al., 2022; Jacobs et al., 2022). Ohlendorf et al. (2021) provide a meta-analysis of literature on the distributional impacts of carbon pricing. They focus on the impact of modelling choices on distributional impact estimates. A broader discussion of their distributional impact and design can be found in Wang et al. (2016).

Microsimulation analyses have also been used to undertake distributional assessments of other environmental policies such as tradable emissions permits (Waduda et al., 2008), taxes on methane emissions from cattle (Hynes et al., 2009) and taxes on nitrogen emissions (Berntsen et al., 2003). Cervigni et al. (2013) have analysed the distributional impact of wider low-carbon economic development policies. A few specialized models were developed to simulate the distributional impact of carbon taxation paired with subsidies for energy efficiency (Giraudet et al. 2021; Bourgeois et al., 2021) and electric vehicles (Ravigné et al., 2022). Others used specialized datasets to assess carbon footprints (Lévay et al., 2021), improve on the granularity vehicle fleets information (Jacobs et al., 2022), or model other revenue recycling mechanisms (Renner et al., 2018).

As inflation rates increase, microsimulation models are again being developed to consider the distributional and welfare impact of price changes (Sologon et al., 2022b, Albacete et al., 2022; Curci et al., 2022; Maier and Ricci, 2024; Amores et al., 2023b), sometimes jointly with carbon taxation (Immervoll et al., 2023).

Existing scalable microsimulation models generally focused on price changes due to indirect taxation (Akoğuz et al., 2020; Amores et al., 2023a), inflation (Sologon et al., 2022b; Menyhért, 2022) or environmental taxation and revenue recycling (Feindt et al., 2021; Steckel et al., 2021; Missbach et al., 2024). These models often ignore the interplay between VAT, ad valorem taxes, excise taxes and environmental taxes, and are limited to simple revenue recycling schemes, or do not estimate behavioural responses. We are only aware of one model that considers a carbon price, revenue recycling, and behaviour (Vandyck et al., 2021). Vandyck et al. (2021) model is foremost a general equilibrium (GE) model, but includes a top-down link to a microsimulation model. Another multi-country GE model linked to a microsimulation model is developed in Chepeliev et al. (2021), but only includes revenue recycling as lump-sum transfer to regional representative households. These models are designed to incorporate adjustments made by producers, and generally provide less granular results regarding differential impacts of price changes across households and household behaviour.

While there is a rich literature analysing the distributional impacts of inflation, indirect taxation, environmental taxation and revenue recycling, and behavioural responses to price changes, the majority of these analysis are disjoint or/and cover a single country. There is a lack of scalable models that are able to simulate the impact of all these forces and their interactions within a common framework. The PRICES model closes this gap by integrating multiple modules into a common modelling framework that utilizes standardized datasets available in many developed and developing countries.

### 3. Methodology and Data

This section describes the methodological approach and dataset used in the framework described above.

*3.1. Data*

The dataset is constructed from two main data sources, the World Input Output Database (WIOD)[1] and a Household Budget Survey (HBS). When sophisticated

---

[1] Because Pakistan is not included as a country in the WIOD, we assume that the technology and import structure used in Pakistan resembles that of India. This is a strong assumption given

mitigation policies are simulated, a third dataset provides detailed information on incomes, taxes and benefits.

HBS datasets contain detailed information on household expenditure by item and information on household demographic and socioeconomic characteristics and income. The application in this paper only utilise the Pakistan HBS for 2018. Unlike the HBS for most countries, the Pakistan HBS does not include an income variable and so we rank households on total expenditure. For the sake of generality of the model description, we refer to income in the model description below. The Pakistan HBS does not record alcohol purchases or childcare expenditure by households.

In applications that require estimates of inter-industry linkages, we use the 2016 WIOD data and its environmental extension reflecting industry-level $CO_2$ emissions (European Commission, 2021; Corsatea, et al., 2019). The WIOD map monetary flows across 56 industries in 44 countries. The use of Multi-regional Input-Output (MRIO) models reflects the state of the art in the estimation of GHG emissions associated to household consumption (Feindt et al., 2021; Steckel et al., 2021). The PRICES framework allows for two approaches to computing the industry-level $CO_2$ emissions. A first approach is to use an emission vector provided Corsatea et al. (2019). This emission vector has one non-negative entry for each industry in each country, and includes process-based and fugitive emissions. A second approach is to compute the $CO_2$ emissions emitted by energy industries in each country and to trace energy use across industries. This approach allows focusing on energy-related emissions only. In this application, we do not utilize the IO component of the model.

---

that India and Pakistan differ substantially in their population size and the size of the economy. The import share of India and Pakistan as a share of GDP is however relatively low at 20.9% and 16% (World Bank, 2024). As argued by Owen (2017), differences in import shares for a given country across different MRIO affect results relatively little. Despite these limitations, we argue that India's Technology and Import structure provides a reasonable approximation for Pakistan, considering the countries available in the WIOD and the associated data limitations.

Simulating sophisticated mitigation measures requires information on income by source (market, investment, capital), benefits and taxes. Specialized surveys, such as the EU Survey on Income and Living Conditions (SILC), contain this information or can be used to produce this information using tax benefit simulation models (e.g. EUROMOD). Additionally, these datasets contain information on household socioeconomic and demographic characteristics that can be used to link them to HBS datasets. We discuss these approaches to linking both datasets in section 3.5. We do not utilize such a dataset in this case study.

*3.2. Theoretical model*

In this paper, we describe a framework that can simulate price changes due to

- Inflation
- Indirect taxation
- Carbon prices

For the purposes of this paper, we test the framework considering the distributional impact of inflation and plan to develop other analyses in due course in relation to indirect taxation and carbon taxation.

The structure of the microsimulation modelling framework is described in Figure 1. The light blue boxes represent the datasets used. The mid-shade blue boxes represent the components implemented in the current version of the model, and the grey boxes represent the components that are not yet implemented in the current version. Finally, the dark blue boxes represent the outputs of the model.

The PRICES model is similar to an indirect tax model, containing input expenditure data, a policy calculator and the consumption behavioural response. At its core is the ability to incorporate mechanism that influence price in the three dimensions; inflation, indirect taxation, and carbon pricing. Starting at the top left of Figure 1, changes in the Consumer Price Index (*Price Change*) and tax rates (*Fiscal Policy*) are sourced from national statistical offices and are determined outside the model. As these initial price changes are not determined within the model, the boxes *Fiscal Policy* and *Price Change* are separated from the other components. The other components of the same colour are determined within the model. For example, an indirect tax policy component uses information on tax rates to compute producer prices and to convert ad valorem and excise taxes into rates, and computes household-level *Tax Payments* using *HBS data*.

Price changes can also be feed through the Input Output model to adjust the composition of a monetary unit of industry output. The Input-Output model is constructed using an *Input Output Database*. The Input-Output model is primarily used to compute price changes due to carbon pricing. The approach to computing price changes due to carbon pricing is described in section 3.3.

For specific sectors with high emission abatement potential, such as the public transport (*Green Public Transport*) or electricity generation (*Green Electricity Generation*) sector, we plan separate model components. Additionally, we plan a component that accounts for the impact of price changes on the structure of the economy (*sectoral changes*) and employment. The impact of employment changes on income can be modelled using an Income Generation Model., which uses *income data* as input and computes new income levels (e.g. Sologon at al. 2020). These changes are not yet modelled in the current version of PRICES but will be incorporated in future versions. Approaches to incorporating these changes are discussed in section 3.2.1.

Once we computed price changes for different expenditure categories, we utilize information on household expenditure from *HBS data* to compute household specific *tax payments* on the household level. Summing tax payments across households gives revenue available to fund *mitigation policies* through *revenue recycling*. Information on direct tax liabilities and benefit payments from the *income data* are used to expand on the mitigation options simulated.

To account for household behavioural responses to price and income changes, the PRICES model includes a simple demand system, described in section 3.5. Price changes influence consumption behaviour via own and cross price effects (*intensive margin*), while *revenue recycling* and *mitigation policies* influence consumption behaviour via income effects (*income behavioural response*). The parameters estimated from the demand system form the basis to estimate the *welfare* impacts and changes in *emissions*.

Lastly, price changes also induce households to invest into new capital, such as electric vehicles, efficient and low-carbon heating systems, and solar panels. Modelling these changes along the *extensive margin* however requires specialised models discussed briefly in section 3.2.1.

**Figure 1.        Structure of a Price based Microsimulation Model**

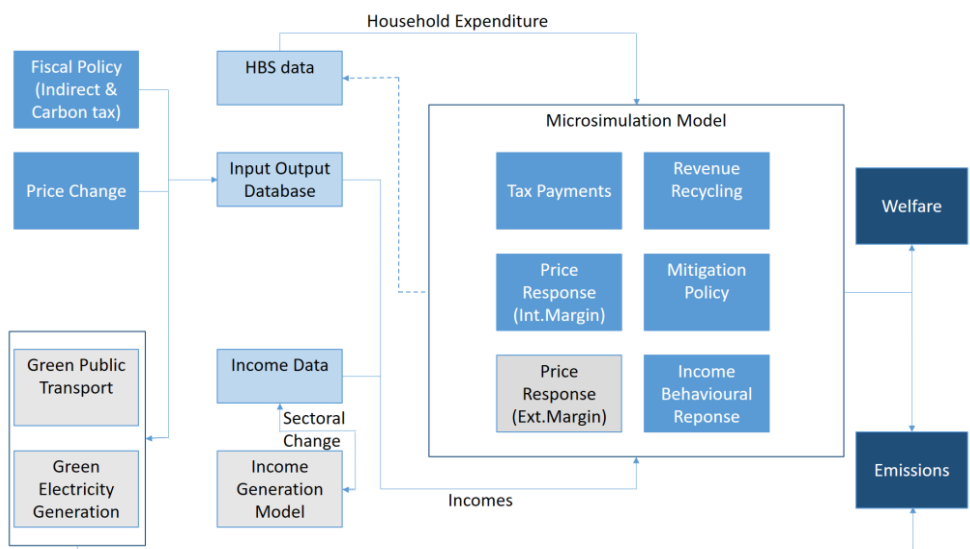

### *3.2.1. Future extensions*

Figure 1 describes four components that are not yet implemented in the current version of the PRICES model, but that are planned as part of an updated version. The first two components concern Green Public Transport and Electricity Generation. The third component concerns employment and income effects of price related policies. The last component relates to innovation adoption by households.

As part of their commitments to reduce Greenhouse Gas emissions, governments are investing in Green Public Transport and Green electricity generation. Large-scale infrastructure investments will affect the carbon intensity of household consumption by changing the Input-Output structure of the public transportation and electricity generation industry. This can be implemented through calibration of the (energy) input structure of the transportation and electricity industry.

Price changes also affect employment and incomes through general equilibrium (GE) effects. Changes in employment and incomes have distributional implications, and were found to reduce the regressivity of carbon pricing (Matcalf, 2023; Rausch and Schwarz, 2016). Commonly, distributional impact of GE effects are assess through models integrating household heterogeneity into macroeconomic models (see for example Rausch and Schwarz (2016), Vandyck et al. (2021), and Antosiewicz et al. (2022)). Bourguignon et al. (2008) and Cockburn et al. (2014) review different approaches. The approach taken in the PRICES model will rely on an Income Generation Model (IGM) developed by Sologon et al. (2020) (see also Sologon et al. (2023) for a review). The IGM contains a parametric representation of the links between different sources of household income and individual and household characteristics, complemented with non-parametric reweighting techniques to account for demographic profiles. The IGM allows accounting for endogenous labour supply adjustments and can be calibrated or reweighting to exogenous changes in the structure of the economy (see for example O'Donoghue et al. (2020) and Sologon et al. (2022a)).

As prices change, the adoption of low-emission vehicles and heating systems, and domestic renewable electricity sources, such as solar panels, will likely become increasingly attractive. Modelling the adoption of new technologies requires specialized models, such as the threshold model (David, 1969; Zilberman et al., 2012), the diffusion of innovation model (Rogers, 2003), or the theory of planned behaviour (Ajzen, 1991). Using these theories and specialized surveys, the approach is to estimate discrete choice models for each technology and to adjust the energy use of households as a function of the technology used.

*3.3. Environmentally Extended Input-Output Model*

Modelling the impact of energy price changes on households' cost of living incorporates both a direct impact on the price of energy consumed by households, and an indirect impact associated with price changes of inputs used in the production of other goods and services consumed by households[2]. A change in the price of inputs used in the production process impacts the producer price of goods and services. This price increase is (partially) passed through to consumers. Here, we focus on energy price changes due to carbon pricing.

In order to capture the indirect effect of producer price changes and carbon taxes, the transmission of price changes through the economy to the household sector is modelled using an input-output (IO) table. IO modelling, initially developed by Leontief (1951), is discussed extensively in Miller and Blair (2009). O'Donoghue (1997) for Ireland, Gay and Proops (1993) in the UK,

---

[2] Indirect emissions can be divided into emissions produced by imports which are typically not taxable and emissions produced by domestically produced goods and services which are likely to be taxable. Similarly direct emissions can be divided into emissions from purchased energy and own produced energy such as harvested firewood etc.

Casler and Rafiqui (1993) provide early example of distributional impact assessments of carbon taxation using IO models. Sager (2019) and Feindt et al. (2021) provide state-of-the-art distributional impact analysis of carbon taxation using Multi-regional IO (MRIO) models.

IO and MRIO models can be complement with environmental extensions to trace the environment impact of production process through the global supply chain, allowing the construction of an Environmentally Extended-MRIO (EE-MRIO). Environmental extensions are vectors of emissions or resource use associated to the production of each sector in each region. In the case of carbon emissions, EE-MRIO models link products to indirect carbon emissions embedded in the production process of goods and services. Kitzes (2013) introduces environmentally extended Input-Output analysis. Minx et al. (2009) provide an overview of applications to the estimation of carbon footprints.

The central equation of an IO model is the Leontief inverse matrix $(I - A)^{-1}$, where $I$ is the identity matrix and $A$ is the technology matrix. The Leontief inverse gives the direct and indirect inter industry requirements for the economy:

$$x = (I - A)^{-1}.d \quad (1)$$

Where $d$ is a vector of final demand.

Transforming an IO model into an EE-IO requires a carbon intensity vector, capturing carbon emissions emitted by the industry in the production of a monetary unit of its output. Multiplying the Leontief inverse with the carbon intensity vector, we get a vector of the carbon intensity of each monetary unit of industry output ($E_{ind}$), accounting for emissions released by the industry and by all downstream industries. Using bridging matrices described in section 3.3.2., we can translate the carbon emissions associated to industry outputs into indirect emissions associated to products consumed by households ($E_{indHH}$).

To compute total household level emissions, we combined information on household fuel consumption with the carbon intensity of each fuel to create a vector of the household's direct carbon emissions ($E_{dirHH}$). The sum of direct and indirect emissions gives households' total carbon emissions associated to their consumption ($E_{HH}$):

$$E_{HH} = E_{dirHH} + E_{indHH} \qquad (2)$$

We provide a description of the IO methodology and its environmental extension in the appendix.

### 3.3.1. Selection of the Multi-Regional Input Output model

Multiple MRIO models and databases exist, with differences in the level of sectorial disaggregation, geographical coverage, their environmental extensions, and accessibility. Commonly used databases include the GTAP database, Eora, EXIOBASE and WIOD. Other databases include the EMERGING database (Huo et al., 2022), focusing on emerging economies, Asian International Input–Output Tables (Meng et al., 2013), OECD–WTO database on TiVA (OECD and WTO, 2013), and the OECD Inter-Country Input Output Database (OECD, 2023). Tukker et al. (2015) and Wiedmann et al. (2011) review different databases, and Tukker and Dietzenbacher (2013) and Steen-Olsen et al. (2014) discuss how the use of different databases may affect modelling results. Differences might arise from multiple sources, including the level of industry aggregation, the source of the emission data, and modelling assumptions (Kymn, 1990; Owen et al., 2014). Timmer et al. (2015) compare the features of WIOD, GTAP and EXIOBASE. Owen et al. (2014) compare carbon footprint estimates produced by GTAP and WIOD, and show that both MRIO databases produce comparable results (Owen et al., 2014), though differences may arise for some countries (Arto et al., 2014). We are not aware of more up-to-date comparisons between the features of major MRIO databases.

Overall, researchers should evaluate which MRIO best suits the purpose of their study. In modelling of carbon emissions, the level of aggregation of the energy sector and energy consumption categories is relevant for the estimates, and may guide the selection of the MRIO database. The level of aggregation of the energy sector is directly relevant for the IO analysis and indirect emission estimates ($\boldsymbol{E}_{indHH}$). For example, GTAP includes two energy sectors, electricity and gas, while WIOD combines both sectors. Environmental extensions also differ across databases. GTAP 7 uses CO2 emissions derived from IEA energy data and covers energy related emissions only, whereas WIOD utilizes NAMEA data and covers fugitive and process related emissions (Owen et al., 2014). EXIOBASE and EORA are specialized for environmental impact accounting, and provide more detailed and diverse emissions vectors.

The level of aggregation of the energy consumption category is relevant for the direct emission estimates ($\boldsymbol{E}_{dirHH}$). Direct emissions are derived directly from the expenditure levels of energy products recorded in expenditure surveys. Some MRIO databases provide information on energy products' carbon intensities per monetary unit of expenditure[3]. Researchers using WIOD often explicitly model direct emissions from fuel combustion by sourcing energy prices and carbon intensity factors for different fuel types (see for instance Sager (2019) and Immervoll et al. (2023)). This approach is sensible when the level of aggregation of the energy products in the MRIO is high (e.g. only one category for domestic energy), or energy products consumed by households are missing in the MRIO. This approach is taken in the PRICES model as WIOD and its EE do not include household energy products. This is motivation by the importance of direct emissions in total household emissions and by the requirement to differentiate between the carbon intensity of different fuels.

---

[3] For instance, GTAP provides users with a vector of household emissions by energy product, whereas WIOD only includes industry level emissions. EXIOBASE and provide more detailed and diverse emissions vectors by product.

Another important consideration is the treatment of the raw data and its proximity to official statistics. WIOD prioritised proximity to official statistics (Arto et al., 2014), whereas GTAP priorities the quality of trade related information. Tukker et al. (2017) show that the main uncertainty in carbon footprint estimates from MRIO analysis results from the environmental extension used. Eora provides measures of uncertainty for the MRIO entries. This is particularly useful to specialized audiences as it enhances transparency but may complicate their use for novice users. EORA compilation is highly automatized and therefore can cover every country, but lacks the manual checking of other MRIO databases, and relies more heavily on judgment by expert users.

Other important consideration in the selection of MRIO include the level of spatial and sectorial aggregation, temporal availability, and the availability of environmental extensions[4].

In the development of PRICES, data availability, usability[5], proximity to official statistics, documentation, and continuation were important considerations. WIOD is freely available and the development of the latest environmental and socio-economic satellite accounts is housed by the European Commission's Joint Research Centre (JRC). GTAP, on the other hand, requires cooperation with its managing team (Lenzen, 2011). Versions of EXIOBASE and Eora are also freely available to academic users. By our understanding,

---

[4] The level of spatial and sectorial disaggregation and temporal availability may also guide the MRIO selection. Relative to WIOD, that includes 56 sectors and 49 countries (including a category for the rest of the world), Eora has higher sectorial and country resolution (Lenzen et al, 2011), and might be particularly interesting for researching modelling the land, energy, and. water use, and emissions from agriculture. Eora covers all countries but relies heavily on imputation methods, which may compromise reliability of the data. Compared to WIOD, EXIOBASE 3.6 provides similar levels of country aggregation and higher levels of industry aggregation. It is available on a product-by-product and industry-by-industry level, and feature higher levels of emission and resource categories. Like WIOD, Eora and EXIOBASE 3 are available for multiple years, with the most recently tables available in Eora and EXIOBASE.
[5] Comparing WIOD to GTAP, Arto et al (2014) point towards the need to adjust the factor and input structure of the gas, electricity and water supply industry for key economies such as China and EU countries in GTAP.

EXIOBASE and Eora are supported by specific research projects. At the time of development, our evaluation was therefore that WIOD is most likely to be a long running, freely available MRIO database. Development of WIOD has however been discontinued since. For an updated version of the model, we are considering moving to the OECD Inter-Country IO (ICIO) tables or EXIOBASE 3. Other reasons to select WIOD include that it requires lower levels of expert knowledge relative to the other MRIO models available, and the availability of all underlying Input-Output and Supply and Use Tables.

*3.3.2. Matching WIOD and HBS*

To compute household's carbon footprints, we combine information from WIOD and HBS. HBS reports expenditure across consumption purposes (COICOP). WIOD reports inter-industry flows and final consumption by industry classification (ISIC rev. 4 or NACE rev. 2). To translate between consumption purpose and industry classifications, we use bridging matrices (Cai and Vandyck, 2020)[6]. A bridging matrix maps the use of a product to satisfy a consumption purpose, so that the $k^{th}$ element of matrix $B = [b_{kj}]$ represents the use share of industry product j for consumption purpose k. Industry products can then be translated into industry output. The integration of HBS data into multi-sectoral models is described in Mongelli et al. (2010) and Cazcarro et al. (2022). We adapt the approach taken by Mongelli et al. (2010) and follow four steps:

1) Transform from consumer product (COICOP) to Industry product (CPA (Classification of products by activity)) using the bridging matrix by Cai and Vandyck (2020)[7].

---

[6] Pakistan is not included in the bridging matrices supplied by Cai and Vandyck (2020). We aggregate the bridging matrices so that the final result is a generalized bridging matrix. We use the generalized bridging matrix as approximation for a Pakistan bridging matrix.

[7] Cazcarro et al. (2022) provide improved bridging matrices. Future versions of PRICES will rely on the bridging matrices supplied by Cazcarro et al. (2022).

2) Match Budget shares to CPA categories by aggregating COICOP categories expenditure categories and calculating the weighted sum of CPA contributions to expenditure categories.

3) Match CPA categories to WIOD using national supply tables to calculate CPA input per industry output using the Fixed Product Sales Structure Assumption[8].

4) Assign the relative contribution of each sector in the country to the appropriate budget shares.

The approach described in Mongelli et al. (2010) has been improved in Cazcarro et al. (2022). In future version of the PRICES model, we will adopt the approach described in Cazcarro et al. (2022). A description of the steps suggested by Cazcarro et al. (2022) can be found in the appendix.

*3.4. Imputation of expenditure patterns into income datasets*

Modelling prices and price related policy requires expenditure data, typically included in HBS. Expenditure data is not typically available in income surveys used in tax-benefit models used to compute direct taxes and transfers, such as EUROMOD (Sutherland and Figari, 2013). Conversely, income surveys do not include detailed information on expenditures. Estimating the net distributional impacts of price changes and diverse mitigation measures requires information on household's consumption patterns, employment situation, incomes, and demographic characteristics. We therefore use datasets with rich data on incomes, such as the EU-SILC, and impute expenditure and expenditure patterns from the HBS into income datasets. Where possible, we use tax-benefit simulation models, such as EUROMOD, to compute households' disposable incomes, including tax liabilities and benefit payments, and impute expenditure patterns into the datasets produced by tax-benefit simulation models.

---

[8] WIOD does not supply a national supply table for Pakistan. We construct an EU wide supply table by summing all single EU country supply tables. This EU-wide supply table serves as proxy for supply tables of countries for which no data is available. This reduces issues relating to country-specialization along industry supply chain and ensures that the composition of a product reflects the majority or all its inputs, rather than only those produced within a specific country.

Different techniques to combine datasets are available. They include the use of parametrically estimated Engel curves, non-parametric estimation, minimum distance matching techniques, or combinations of both (for example, Akoguz et al., 2020). Comparison of the different techniques are provided in Akoguz et al. (2020) and Decoster et al. (2020). While matching techniques may appear favourable theoretically[9], Decoster et al. (2020) show that empirically parametric and non-parametric Engel curve estimation yield the best results at the mean. Minimum distance methods produce better distributions within and between variables (Decoster et al., 2020). The best choice of imputation method therefore depends on the application of interest and computation constraints. Particularly when datasets are large, minimum distance methods are computationally costly. Akoguz et al. (2020) use a combination of minimum distance matching and parametric Engel curves to obtain the best estimate on average and introduce granular information on expenditure products.

The PRICES model uses parametrically estimated Engel curves to impute expenditure patterns into datasets containing information on income sources and tax payments. We follow a three-step procedure. In a first step, we impute total expenditure as a function of disposable income and household characteristics. In a second step, we collapse all items in the HBS into 19 expenditure categories and impute the likelihood of positive expenditures using a logit model to account for zero expenditures (e.g. for motor fuels). The domestic fuel and motor fuel categories are further sub-divided; distinguishing between two motor fuels (diesel and petrol) and five heating fuels (liquid fuels, gas, coal, fire wood, district heating). In a third step, we impute the conditional budget share of each category. Sub-categories for motor and domestic fuels are imputed following the same procedure. Our approach broadly follows that described in De Agostini et al. (2017).

---

[9] They do not rely on theoretical assumption or specifications of functional forms and avoid problems relating to zero and infrequent expenditures discussed below, and they do not require that expenditures are grouped.

First, we estimate total expenditure as a function of household disposable income and demographic characteristics available in both datasets:

$$\ln c^h = \alpha + \beta \ln y^h + \delta q^h + \varepsilon^h \qquad (3)$$

Where $c^h$ is total consumption expenditure of household $h$, $y$ is household disposable income, $q$ is a vector of demographic characteristics[10] and $\varepsilon$ is the error term. We generate a normally distributed error term, reproducing the mean and variance of the error term in the HBS.

Next, we estimate the likelihood of having positive expenditure for each expenditure category:

$$\Pr(d_i^h = 1) = \emptyset(\alpha_i + \beta_i \ln c^h + \gamma_i (ln\ c^h)^2 + \delta q^h + \varepsilon^h) \qquad (4)$$

We then rank households according to this likelihood and assign positive expenditure to the highest ranked households until the share of households with positive expenditure in the HBS is replicated. A drawback with this approach is that it is unlikely to replicate expenditures for groups where zero expenditure is very common. An alternative approach is to draw from a uniform distribution between zero and one and assign positive expenditures if the drawn number exceeds the fitted probability. This ensure that some households with low probabilities are also assigned positive expenditures.

---

[10] Demographic characteristics can vary depending on the data availability and structure. The base model includes: household size, the number of earners, occupation of the household head, age of the household head, gender of the household head, marital status of the household head, the level of education of the household head, number of children aged 0-5, number of children aged 6-13, number of children aged 16-24, number of retired persons, number of adults, household type above (below) median income single, couple, couple with children, single parent), and a set of binary variables indicating whether the household head is self-employed, an employee, a student, a blue-collar worker. For subcomponent, this also includes the level of expenditure of the component (e.g. for diesel this includes expenditure on motor fuels).

Next, we estimate Engel curves for each expenditure category, conditional on having positive expenditure, where $w_i^h = \frac{e_i^h}{c^h}$:

$$w_i^h = \alpha_i + \beta_i \ln c^h + \gamma_i (\ln c^h)^2 + \delta q^h + \varepsilon^h \quad if \ d_i^h = 1 \tag{5}$$

Where $w_i$ is the budget share allocated to good $i$.

For each equation, the estimated parameters are used to impute total expenditure, the presence of expenditure, and budget shares into the income dataset using variables contained in both datasets. Before estimating the equations described above, we calibrate disposable income in the HBS to reflect the mean and standard deviation of disposable income in the income dataset. The mean and standard deviation are calculated without extreme values, which are determined using the Chauvenet's criterion, following De Agostini et al. (2017). Finally, the sum of imputed budget shares is adjusted to equal 1.

To validate the imputation method, we follow two steps. In a first step, for each equation, we compare the estimated coefficients with measures from the literature to the estimated coefficients on important variables (the log of income and it square in equation (3) and the log of expenditure and its square in equation (4) and (5)) (e.g. O'Donoghue et al., 2004). In a second step, we compare the distribution of the imputed variables to the distribution of the variables in the source dataset.

A common approach to validation of microsimulation models is to compare the simulated results to official aggregate statistics and to investigate the source of differences between simulated and official totals (O'Donoghue, 2014). With expenditure data, validating the imputed totals against official statistics, such as National Accounts, is problematic because expenditure on certain items are often misreported, particularly items such as alcohol and tobacco (Atkinson and Micklewright, 1983; Banks and Johnson, 1997). In principal, simulated totals can be reweighted to reflect official aggregates. This however requires the assumption that the misreporting and other sources of differences between simulated and official aggregates, due to differences in survey weights used, and uprating and imputation techniques (Immervoll and O'Donoghue, 2009), are distributed equally across households. In the case of expenditure data, if reweighting is applied to each expenditure item and some households are more likely to misreport than others are, this may introduce additional bias in the structure of households' budget. Reweighting should be used with caution as it can substantially alter simulation outcomes (Myck and Najsztub, 2014).

*3.5.Behavioural Estimates*

In order to model behaviour, a demand system is required that relates the consumption of a particular good to the price of the good, the prices of other goods, the income of the household and the characteristics of the household. See Deaton and Muellbauer (1980b) for an introduction to this field.

The objective of a demand system is to model households' expenditure patterns on a group of related items, in order to obtain estimates of price and income elasticities and to estimate consumer welfare. This has been popular since Stone's (1954) linear expenditure system (LES). The dependent variable is typically the expenditure share.

Two of the most popular methods are the translog system of Christensen et al. (1975) and the Deaton and Muellbauer (1980a) almost ideal demand system (AIDS), with the latter extended by Banks et al. (1997) to include a quadratic expenditure term (QUAIDS). Estimating a demand system such as QUAIDS requires sufficient price variability to be able to identify the parameters within the system. Frequently however there are not enough data, typically drawn from a number of different years of Household Budget Surveys, to be able to do this.

Therefore in this section a simpler method is described, drawing upon Stone's Linear Expenditure System and described in Creedy (1998).

Rather than estimating a system of demand equations, Creedy (1998) relies on a method due to Frisch (1959) that describes own and cross-price elasticities in terms of total expenditure elasticities ($\eta_i$), budget shares ($w_i$) and the Frisch marginal utility of income parameter ($\xi$) for directly additive utility functions [11]. Own-and cross-price elasticities can be described as follows:

$$\eta_{ij} = -\eta_i w_j \left(1 + \frac{\eta_j}{\xi}\right) + \frac{\eta_i \delta_{ij}}{\xi}, \qquad (6)$$

where $\delta_{ij} = 1$ if i = j, 0 otherwise.

The total expenditure elasticity ($\eta_i$) can be defined:

$$\eta_i = 1 + \frac{dw_i}{dC}\frac{C}{w_i} = 1 + (\beta_i + 2\gamma_i \ln C)/w_i \qquad (7)$$

where $C$ is total consumption expenditure. We estimate $\beta_i$ and $\gamma_i$ using OLS regression and the same specification as in equation (5). Further, to allow for differences in behaviour across population groups, we calculate $w_i$ and $C$ for 10 population groups[12]. We omitted subscripts for population groups in equation (7) to improve readability.

The Frisch parameter ($\xi$), can be defined as the elasticity with respect to total per capita nominal consumption spending of the marginal utility of the last dollar optimally spent (see Powell et al., 1974). In absence of price and quantity data, it is impossible to estimate the Frisch parameter directly and it is necessary to rely on extraneous information. Deaton (1974) provide a review of Frisch parameters. Lluch et al. (1977) empirically estimate the relationship between per capita GNP and the Frisch parameter. This model bas been used to estimate Frisch parameters for multiple countries (Creedy, 2002; Clements et al., 2020; Clements at al., 2022). Lahiri et al. (2000) have estimated a cross-country equation based on 1995 prices relating $-1/\xi = 0.485829 + 0.104019*\ln(\text{GDP pc})$. Estimates for USA, Japan, EU and Australia are respectively -1.53, -1.41, -1.61 and -1.71. A method due to Creedy (2001) (adapted using the exchange rate parameter ER) elaborated on the Lluch et al. (1977) model as follows:

$$\ln(-\xi) = \phi - \alpha \ln(C/ER + \vartheta) \qquad (8)$$

where the parameters $\phi$, $\alpha$ and $\vartheta$ are ad hoc parameters (here respectively 9.2, 0.973, 7000). Note consumption, in this case, is expressed as consumption per capita per month. In the PRICES model, the maximum value of the Frisch parameter has been set at -1.3.

---

[11] See Creedy (1998) for more details.

[12] The ten population groups represent five population groups for two income groups. The population groups are singles, single person with children, couple without children, couples with children, and other households. The two income groups are below and above median income.

The LES has two assumptions. Firstly, the LES is based on the Stone-Geary Utility function, which assumes additive utility functions, i.e. that the utility derived from the consumption of one product is independent of the consumption of other products. This excludes complementary goods and inferior goods. Powell (1974) and Creedy and Van De Ven (1997) argue that when the LES is estimated on aggregated expenditure categories, complementary goods likely fall into the same expenditure category, making the lack of complements and inferior goods acceptable to overcome data limitations. Second, the LES assumes proportionality between income and price elasticities. Clements (2019) find empirical support for such proportionality.

Table 1 reports budget and price elasticities derived from the LES system using our data. For purchased goods, budget elasticities are lower for necessities (Food, fuel, clothing), and for tobacco and recreation, as expenditure on these goods varies less with income compared to expenditure on other goods. Health and communications also have budget elasticities of less than 1, while most other categories have a budget elasticity of about 1. Private education expenditure and durables have a budget elasticity well above 1, indicating that these expenditures are disproportionally concentrated among households with the highest expenditure.

Given the direct relationship between budget and price elasticities, imputed own-price elasticities have a high correlation with budget elasticities. Necessities and other goods with a low budget elasticity are relative price insensitive, while goods such as durables, education and household services are almost perfectly price-elastic. Cross-price elasticities are not reported, but are small relative to own price elasticities.

**Table 1.    Budget and Own Price Elasticities**

|  | Budget | Price |
|---|---|---|
| Food and Non-alcoholic beverages | 0.760 | -0.626 |
| Alcoholic beverages* | . | . |
| Tobacco | 0.455 | -0.275 |
| Clothing and footwear | 0.718 | -0.469 |
| Domestic Energy | 0.165 | -0.099 |
| Electricity | 0.477 | -0.285 |
| Rents | 1.020 | -0.617 |
| Household services | 1.784 | -1.060 |
| Health | 0.833 | -0.514 |
| Private transport | 1.120 | -0.671 |
| Public transport | 1.071 | -0.650 |
| Communication | 0.845 | -0.512 |
| Recreation and culture | 0.284 | -0.171 |
| Education | 1.529 | -0.919 |
| Restaurants and hotels | 1.103 | -0.664 |
| Other goods and services | 1.068 | -0.694 |

| Childcare costs* | . | . |
| Motor fuels | 0.723 | -0.450 |
| Durables | 1.800 | -1.068 |

*Alcoholic beverages and Childcare costs are not included in the Pakistan HBS.

*3.6. Welfare Concepts*

The primary approach to measuring welfare in the PRICES model is to obtain a money metric of the change in welfare following a change in price, the compensating variation (CV). CV is the monetary compensation required to maintain the utility of a household at the same (pre-price change) level after a price change. Positive amounts of CV indicating a welfare loss. To compute CV, a utility function is required. As in the case of Creedy (2001) and Sologon et al. (2022b), a Stone-Geary LES direct utility function is utilised:

$$U = \prod_i [\pi_i - \gamma_i]^{\phi_i} \tag{9}$$

where $\gamma_i$ are LES parameters known as committed consumption for each good $i$ and $\phi_i$ is the marginal utility from an extra unit of consumption beyond committed consumption of good i, and is $0 \leq \phi_i \leq 1, \sum_i \phi_i = 1$. For convenience, we ignore the subscripts indicating that different parameters are estimated for different demographic (and income) groups.

The concepts $\phi_i$ and $\gamma_i$ can be derived from (9) (see derivations in the appendix). They form the basis of the compensating variation, given by

$$\Delta W_{CV}^h = CV^h = \left[\sum_i p_{1i}\gamma_i^h + \prod_i \left[\frac{p_{1i}}{p_{0i}}\right]^{\phi_i} \left(y_0^h - \sum_i p_{0i}\gamma_i^h\right)\right] - y_0^h \tag{10}$$

Where subscript h indicates the household, $p_{0i}$ and $p_{1i}$ is the price of good i before and after the price change, $y_0^h$ is household income before the price change (see Sologon et al. (2022b) for further detail). As the Pakistan data does not have data on household incomes, we use expenditure instead of income.

## 4. Results I: Expenditure Patterns and Price Changes

In this section we describe the distributional impact of price changes in Pakistan over the period 2020 Q4 to 2022 Q4. The differential impact of inflation across the distribution is driven primarily by the good specific price changes and the budget share of these expenditures across the distribution.

Figure 2 reports the average price change over the two-year period by COICOP expenditure category. Unsurprisingly transport costs have the highest price growth rate at nearly 80%, followed by food and drink sectors at about 50%. Domestic energy fuels experienced a relatively low price growth of about 25%.

**Figure 2.    Expenditure Category specific Price Growth Q4 2020-Q4 2022**

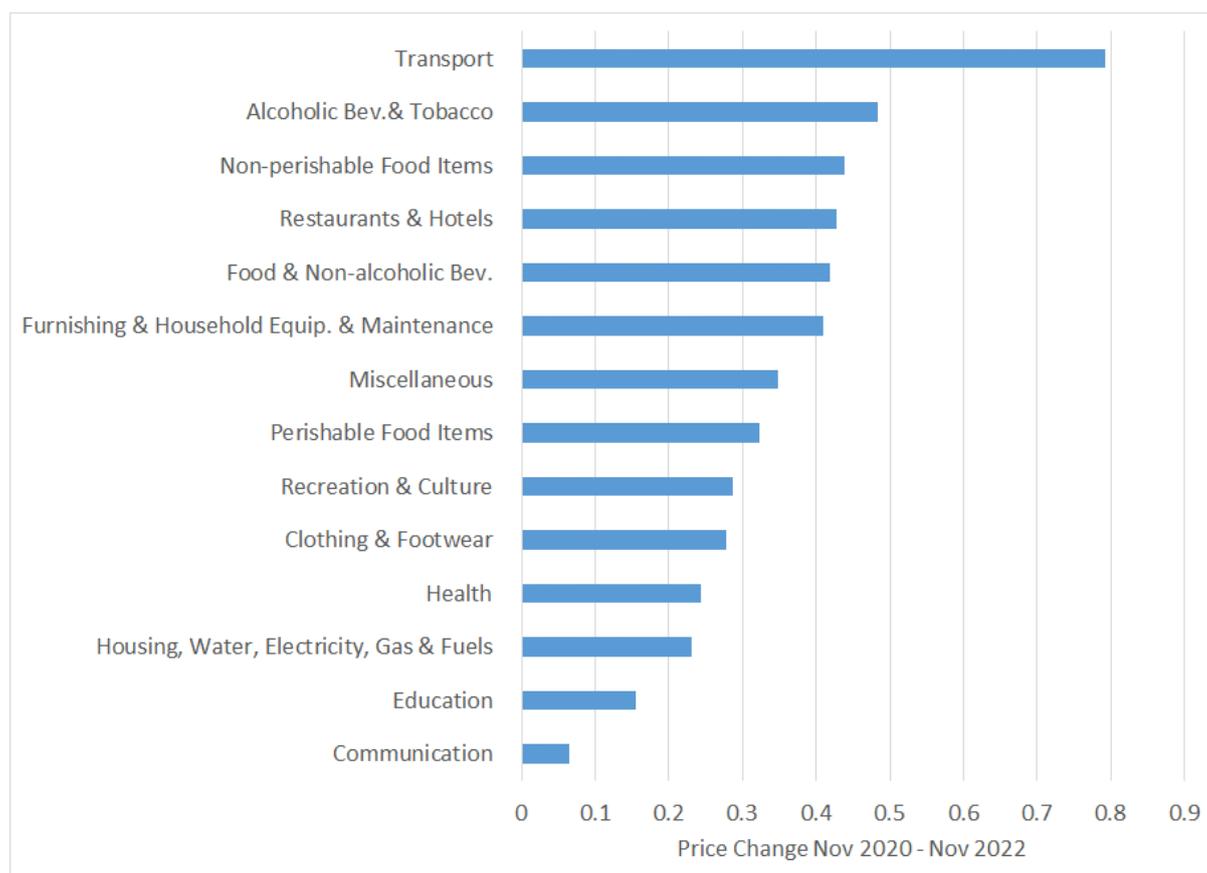

Source: Pakistan Bureau of Statistics Consumer Price Index

Table 2 decomposes inflation into four high-level expenditure groups, covering the main differential inflation rates; food, motor fuels, domestic energy and electricity and other goods and services. In total, the energy budget share is relatively low in Pakistan at about 5%, with the food budget share at over 40% and the remainder allocated to other goods and services. Whilst fuel price inflation is much higher than in other areas, the lower budget share means that the contribution made by fuels to the overall inflation rate is relatively low.

**Table 2.    Drivers of Inflation by High Level Expenditure Groups**

|  | Budget Share | Average Inflation Rate (in %) | Contribution to total inflation (in %) |
|---|---|---|---|
| Food | 0.417 | 42.89 | 17.89 |
| Motor Fuels | 0.047 | 79.27 | 3.74 |
| Domestic Energy and Electricity | 0.007 | 63.65 | 0.44 |
| Other Goods and Services | 0.529 | 36.61 | 19.36 |
| Total | 1.000 | 41.43 | 41.43 |

Table 3 describes the budget share by equivalised expenditure quintile. The budget share for food in the bottom quintile is more than 50% falling to 34% in the top quintile. Purchased domestic energy and electricity have a slightly higher share in the bottom of the distribution. Conversely, the budget shares for motor fuels and other goods and services rise over the distribution. The average expenditure of households in the lowest quantile is 42% of the population average, while those in the highest quantile spend approximately twice as much as the average household.

**Table 3.    Budget Shares of expenditure components across equivalised expenditure quintiles (as a share of total expenditure)**

| Quintile | Food | Motor Fuels | Domestic Energy and Electricity | Other Goods and Services | Relative Expenditure |
|---|---|---|---|---|---|
| 1 | 0.511 | 0.031 | 0.008 | 0.450 | 0.42 |
| 2 | 0.487 | 0.041 | 0.007 | 0.464 | 0.62 |
| 3 | 0.467 | 0.043 | 0.007 | 0.483 | 0.82 |
| 4 | 0.444 | 0.049 | 0.007 | 0.500 | 1.11 |
| 5 | 0.342 | 0.053 | 0.006 | 0.599 | 2.03 |
| Average | 0.450 | 0.044 | 0.007 | 0.499 | 1.00 |

Table 4 shows the budget share by equivalised expenditure quintile before and after inflation. The budget share for food increase for all households expect the richest. Budget shares on goods and services decrease at the bottom three quantiles, but increase for the top two quantiles. Both motor fuel and domestic energy and electricity budget shares fall for all households.

**Table 4. Budget Shares before and after Inflation without behaviour**

| Quintile | Food | | Motor Fuels | | Domestic Energy and Electricity | | Other Goods and Services | |
|---|---|---|---|---|---|---|---|---|
| | Before | After | Before | After | Before | After | Before | After |
| 1 | 0.511 | 0.524 | 0.031 | 0.027 | 0.008 | 0.007 | 0.45 | 0.442 |
| 2 | 0.487 | 0.499 | 0.041 | 0.036 | 0.007 | 0.006 | 0.464 | 0.459 |
| 3 | 0.467 | 0.476 | 0.043 | 0.038 | 0.007 | 0.006 | 0.483 | 0.481 |
| 4 | 0.444 | 0.450 | 0.049 | 0.043 | 0.007 | 0.006 | 0.5 | 0.501 |
| 5 | 0.342 | 0.331 | 0.053 | 0.046 | 0.006 | 0.005 | 0.599 | 0.618 |
| Average | 0.45 | 0.456 | 0.044 | 0.038 | 0.007 | 0.006 | 0.499 | 0.500 |

The distributional impact of inflation across equivalised expenditure quintiles is described in Table 5. The columns in Table 5 show the inflation rate on each expenditure group weighted by its budget share for each equivalised expenditure decile. Reflecting their higher budget share, the distributional pattern is driven by the relative budget shares of food and other goods and services. Food makes a higher contribution to inflation at the bottom of the distribution, while other goods and services and motor fuels make the largest contribution at the top of the distribution. Overall the average inflation rate is higher at the top of the distribution.

**Table 5.   Distributional impact of inflation across equivalised expenditure quintiles**

| Quintile | Food | Motor Fuels | Domestic Energy and Electricity | Other Goods and Services | Average |
|---|---|---|---|---|---|
| 1 | 0.215 | 0.025 | 0.005 | 0.163 | 0.408 |
| 2 | 0.206 | 0.033 | 0.005 | 0.165 | 0.409 |
| 3 | 0.198 | 0.034 | 0.004 | 0.170 | 0.407 |
| 4 | 0.190 | 0.039 | 0.004 | 0.178 | 0.412 |
| 5 | 0.149 | 0.042 | 0.004 | 0.226 | 0.422 |
| Average | 0.192 | 0.035 | 0.005 | 0.181 | 0.411 |

Columns show the composition of the inflation (shown in the last column) in terms of the inflation in each expenditure category across equivalised expenditure quintiles.

*4.3. Welfare Impacts*

We evaluate next how the cost of living was affected by the price increases and the contribution of prices changes towards households' welfare. We measure the impact on household welfare using the compensation variation (CV). CV represents the monetary compensation that households should receive in order to maintain their initial well-being (utility) after the price increases. In Table 7, we express CV relative to total initial expenditure for households along quintiles of household equivalised expenditure in order to approximate the percentage change in the cost of living for households with different means.

**Table 6.   Welfare losses decomposition into price and behavioural adjustment**

| Equivalised Expenditure Quintile | Inflation | Relative CV | Behaviour |
|---|---|---|---|
| 1 | 0.4087 | 0.4053 | -0.0034 |
| 2 | 0.4105 | 0.4068 | -0.0037 |
| 3 | 0.4087 | 0.4060 | -0.0027 |
| 4 | 0.4131 | 0.4123 | -0.0009 |
| 5 | 0.4215 | 0.4197 | -0.0017 |
| Total | 0.4125 | 0.4100 | -0.0025 |

Whereas the relative increase in costs due to inflation captures the increase in expenditure that households face due to price increases given their current consumption pattern, relative CV (welfare losses) captures the relative increase in income that households would need in order to maintain their utility under the new prices. The difference between them represents the adjustment that households do in their consumption behaviour (due to changes in the relative prices between different commodity groups) in order to maintain their utility under the price increases. In other words, how much would the price increase cost households without a behavioural adjustment minus how much it would cost taking into account that households can modify their behaviour.

Overall, it appears that the behavioural response component has very limited effects on welfare. The picture of welfare losses along the distribution of income follows the same distributional pattern of inflation above. The distributional impact of inflation is slightly more progressive once the behavioural component of the compensating variation is accounted for. To understand this results, we must consider the inflation rates of different goods, the own-price elasticities of different goods and the composition of the consumption basket along the income distribution. High inflation rates are recorded for food and transport, which have relatively high own-price elasticities. Food budget shares are substantially higher than transport budget shares. Further, food budget shares are substantially higher for low-income households than for high-income households. Higher food budget shares, paired with high food inflation and large differences in food budget shares between low-and high-income households lead to larger behavioural responses among low-income households than high-income households. These factors together lead to higher overall behavioural responses for low-income households.

# 5. Conclusions

This paper develops the microsimulation framework PRICES (Prices, Revenue recycling, Indirect tax, Carbon, Expenditure micro Simulation model) to simulate the distributional impact of price changes and price-related policies, including indirect and carbon taxes. The framework provides a static incidence analysis of these issues, combining a model to incorporate price and income related behavioural responses, linked with an Input-Output framework to capture value chain transmission of price changes. As a pilot exercise, an analysis of the distributional impact of price changes during the cost of living crisis between 2020 and 2022 was evaluated for Pakistan.

The cost of living crisis was marginally progressive in nature with slightly higher price increases at the top of the distribution than the bottom. This reflects expenditure patterns across the distribution and the good specific price changes. While energy prices increased more in Pakistan than in other countries, the relatively low budget share (particularly of purchased fuels) means that the net impact on welfare due to increases in energy prices is relatively low. The biggest driver of the welfare loss at the bottom was food price inflation, which comprises over half the budget share, while other goods and services were the biggest driver at the top of the distribution. The distributional impact of inflation in other goods and services, paired with the impact of motor fuel inflation, outweighs the distributional impact of food inflation, so that the overall distributional impact of inflation is mildly progressive.

The role of household behaviour in mitigating the welfare loss is limited, on average allowing households to mitigate just overall half a percent of the increase in their cost of living. The overall distributional pattern of inflation is largely unchanged by households' behavioural response, though low-income households adjust their consumption more than high-income households do. Using the estimated expenditure model, we compute the compensating variation; a money metric of household welfare change. This metric shows that on average, households would need to receive a transfer equivalent to 40% of their pre-inflation expenditure to maintain their utility level at pre-inflation levels, in the absence of income changes.

It should be noted that the distribution is constructed using expenditure instead income. This implies that the role of savings is disregarded. Savings however cushion again price shocks. Generally, lower income households have negative or low savings rates. This implies that a price increase may force low-income households to take up loans and push them into debt, particularly when their behavioural response is limited. Therefore, the cost of living crisis is likely more regressive, and regressive overall, if the distribution is constructed using disposable income and savings rates are taken into account (Davies et al., 1984; Poterba, 1991; Grainger and Kolstad, 2010).

Models similar to the PRICES model provided valuable analysis in recent crisis, where people have lost income sources, like during the COVID-19 pandemic lockdowns (O'Donoghue et al., 2020; Sologon et al., 2022a; Lustig et al., 2021; Doorley et al., 2020; Li et al., 2022; Bruckmeier et al., 2021), or experienced rapid price growth, like during the 2021-2023 cost of living crisis (Menyhért, 2022; Curci et al., 2022; Sologon et al., 2022b). Traditionally, these models have been used to assess the distributional impacts of tax-benefit systems (Sutherland and Figari, 2013), including environmental taxes (Feindt et al., 2021; Cornwell and Creedy, 1996) and other indirect taxes (Decoster et al., 2010; 2011). The PRICES model expands on the available tools to assess distributional impacts of price changes by integrating multiple data sources into a unified scalable model.

A central contribution of the PRICES model is that it unifies multiple aspects important for the distributional impact of price changes in a single framework using standardized datasets available across many countries. The PRICES model accounts for differences in consumption patterns along the income distribution and estimates income and price elasticities for different household types and income groups. It includes the calculation of VAT, excise and ad valorem taxes, allowing the model to compute producer prices. It includes an environmentally extended multi-regional input output model to simulate the impact of input price changes due to policies (such as a carbon tax) on producer prices. The explicit modelling of indirect taxes allows the assessment of the joint impact of indirect taxes and carbon taxes on consumer prices. Lastly, the model includes a procedure to impute expenditure patterns into datasets with richer information on households' employment situation, income sources and tax liabilities. The resulting dataset can be used to assess the net distributional impact of price changes and complex mitigation measures, accounting for global supply chains, and price and income responses by households.

Lustig, N., Neidhöfer, G., & Tommasi, M. (2020). Short and long-run distributional impacts of COVID-19 in Latin America (Vol. 96). Tulane University, Department of Economics.

Lydon, R. (2022). Household characteristics, Irish inflation and the cost of living. Central Bank of Ireland Economic Letter, 2022(1).

Madden, D. (1995). An analysis of indirect tax reform in Ireland in the 1980s. Fiscal Studies, 16(1), 18-37.

Maier, S., & Ricci, M. (2024). The redistributive impact of consumption taxation in the EU: Lessons from the post-financial crisis decade. Economic Analysis and Policy, 81, 738-755.

Maitino, M. L., Ravagli, L., & Sciclone, N. (2017). Microreg: a traditional tax-benefit microsimulation model extended to indirect taxes and in-kind transfers. International Journal of Microsimulation, 10(1), 5-38.

Mardones, C., & Mena, C. (2020). Economic, environmental and distributive analysis of the taxes to global and local air pollutants in Chile. Journal of Cleaner Production, 259, 120893.

Mathur, A., & Morris, A. C. (2014). Distributional effects of a carbon tax in broader US fiscal reform. Energy Policy, 66, 326-334.

Meng, B., Zhang, Y., & Inomata, S. (2013). Compilation and applications of IDE-JETRO's international input–output tables. Economic Systems Research, 25(1), 122-142.

Menyhért, B. (2022). The effect of rising energy and consumer prices on household finances, poverty and social exclusion in the EU. Publications Office of the European Union, Luxembourg, DOI, 10, 418422.

Picos-Sánchez, F., & Thomas, A. (2015). A Revenue-neutral Shift from SSC to VAT: Analysis of the Distributional Impact for 12 EU—OECD Countries. FinanzArchiv/Public Finance Analysis, 278-298.

Pigou, A.C. (1920). The Economics of Welfare. Palgrave Macmillan London.

Poltimäe, H., & Võrk, A. (2009). Distributional effects of environmental taxes in Estonia. Estonian Discussions on Economic Policy, 17.

Poterba, J. M. (1991). Tax Policy to Combat Global Warming: On Designing a Carbon Tax. NBER Working Paper, (w3649).

Powell, A.A., (1974). Empirical analytics of demand systems. Mass., Lexington Books.

Rausch, S., Metcalf, G. E., & Reilly, J. M. (2011). Distributional impacts of carbon pricing: A general equilibrium approach with micro-data for households. Energy economics, 33, S20-S33.

Rausch, S. & Schwarz, G. A. (2016). Household heterogeneity, aggregation, and the distributional impacts of environmental taxes. Journal of Public Economics, 138:43–57

Ravigné, E., Ghersi, F., & Nadaud, F. (2022). Is a fair energy transition possible? evidence from the french low-carbon strategy. Ecological Economics, 96:107397.

Redmond, G., Sutherland, H., & Wilson, M. (1998). The arithmetic of tax and social security reform: a user's guide to microsimulation methods and analysis (Vol. 64). Cambridge University Press.

Renner, S., Lay, J., & Greve, H. (2018). Household welfare and co2 emission impacts of energy and carbon taxes in Mexico. Energy Economics, 72:222–235.
53

**Appendix**

**A.1. Methodological Steps**

- Methodological Steps
- Identify potential policy mechanisms such as Carbon Taxes – produce a carbon tax per tCO2
- Source Input-Output Table
- Source a Household Budget Survey
- Classify expenditure categories into adjusted COICOP headings
- Link IO Categories to HBS categories
- Estimate Budget Elasticity Equations – watch out for zeros. Consider appropriate demographic groups
- Use LES system to derive Price Elasticities
- Run Carbon Tax price change through IO analysis
- Generate Direct and Indirect Price changes
- Run resulting Price changes through the Microsimulation Model
- Derive first round distributional impact of carbon tax
- Derive impact of price change on CO2 using price elasticities
- Develop a suite of mitigation measures in HBS
- If tax-benefit, then impute expenditure into Income Survey
- Incidence Analysis of Expenditures by Type
- Calculate First round impact of Environmental Policies and Price Changes
- Calculate Behavioural impact of Environmental Policies and Price Changes
- Code to link expenditure and micro model
- Direct Tax and Social Protection Simulations to mitigate impact of Environmental Policy
- Indirect Tax and Subsidy Simulations to mitigate impact of Environmental Policy
- Calculate Behavioural impact of Net Impact of Environmental Tax and Mitigation Measures

## A.2. Fuel Prices

**Table 7.   Fuel Prices used in Analysis in Local Currency**

| Year | 2018 |
|---|---|
| Diesel per litre | 73.4 |
| Petrol per litre | 87.3 |
| Electricity per kwh | 10.4 |
| Kerosene per litre | 83.6 |
| LPG per litre | 50.2 |
| Coal per kg | 11.3 |

## A.3. Input Output Model

An IO table contains information about sectors of an economy, mapping the flows of inputs from one sector to another or to final demand (that consumed by households, NGOs, governments, or exported, etc.). Output in each sector has two possible uses; it can be used for final demand or as an intermediate input for other sectors. In an $n$ sector economy, final demand for sector $i$'s produce is denoted by $d_i$ and the output of sector $i$ by $x_i$. Intermediate input from sector $i$ into sector $j$ is defined as $a_{ij} x_j$, where the input coefficients $a_{ij}$, are fixed in value. In other words, $a_{ij}$ is the quantity of commodity $i$ that is required as an input to produce a unit of output $j$. Output can therefore be seen as the sum of intermediate inputs and final demand as follows:

$$x_i = \sum_i a_{ij} x_j + d_i \tag{11}$$

or in matrix terminology:

$$x = A.x + d \tag{12}$$

Combining the output coefficients to produce a $(I - A)$ technology matrix and inverting, the Leontief inverse matrix $(I - A)^{-1}$ is produced, which gives the direct and indirect inter industry requirements for the economy:

$$x = (I - A)^{-1}.d \qquad (13)$$

This can be expanded to produce the following

$$x = (I + A + A^2 + \cdots + A^n)\ .d \qquad (14)$$

As $A$ is a non-negative matrix with all elements less than 1, $A^n$ approaches the null matrix as $n$ gets larger, enabling us to get a good approximation to the inverse matrix. It thus expands output per sector into its components of final demand $d, Ad$, the inputs needed to produce the number of units of each output used in the production of a unit of final demand for each good.

If tax $t$ is applied and is passed on in its entirety to final demand, then the tax on goods consumed in final demand is $td$, the tax on the inputs to these goods is $tAd$, the tax on inputs to these is $tA^2d$ and so on. Combining, total tax is

$$x = (I + A + A^2 + \cdots + A^n)\ t.d. = (I - A)^{-1}t.d. \qquad (15)$$

The original IO table contains information on three fuel sector, Mining and quarrying, Manufacturing of coke and refined petroleum products, and Electricity, gas, steam and air conditioning supply. Because of the focus on the differential effect of price changes on individual fuels such as petrol, diesel, gas and other fuels, this component of the IO table is decomposed into its constituent parts.

In this paper we utilise Multi-regional Input-Output (MRIO) tables from the WIOD. MRIO tables extend the Input-Output (IO) methodology introduced by Leontief (1951). MRIO datasets consist of a matrix mapping the monetary flows between n sectors and m regions, $\mathbf{Z} \in \mathbb{R}^{(m \cdot n)(m \cdot n)}$, with single entries $z_{s1,r1}^{s2,r2}$ representing the monetary flows from sector 1 in region 1 into sector 2 in region 2, and a final demand vector $Y \in \mathbb{R}^{(m \cdot n)(m \cdot n)}$. In a $n$ sector economy, the final demand for sector 1 in region 1 from region 2 and sector 2 is denoted by $f_{s1,r1}^{s2,r2}$ and sector 2's output in region 2 is $x_{s2,r2}$. The input coefficient input coefficient from sector 1 in region 1 into sector 2 in region 2 is the given by $a_{s1,r1}^{s2,r2} = z_{s1,r1}^{s2,r2}/x_{s2,r2}$, where $x_{s2,r2} = \sum_s \left( \sum_r z_{s1,r1}^{s,r} + \sum_r z_{s2,r2}^{s,r} \right)$.

The WIOD includes an environmental extension under the form of a vector of carbon emissions associated to the production of each sector in each region, allowing the construction of an Environmentally Extended-MRIO (EE-MRIO). EE-MRIO models link products to indirect carbon emissions embedded in the production process. Kitzes (2013) provides a short introduction to environmentally extended Input-Output analysis. Let $E \in \mathbb{R}^{(1 \cdot n)}$, $E \in \mathbb{R}(1 \cdot n)$ denote the emissions where $E_i$ refers to emissions produced in sector i. Dividing $E$ entry-wise by the corresponding sector's output, $E_{ind}$ gives the level of CO2 emissions per monetary unit of the sector's output vector. This approach however does not allow differentiating between emissions due to energy use and other emissions, such as fugitive emissions or process emissions. In practice, carbon prices are frequently levied on energy commodities can consequently and thus on the inputs purchased from energy industries. We therefore compute the carbon intensity of energy industry inputs. Calculating carbon intensity of the energy industry requires assumptions on the fuel mix used by domestic energy industries. In this version of the model, we approximate energy industries' fuel mix through the average fuel mix across EU energy industries, sourced from UNIDO MINSTAT[13]. Additionally, to account for differential carbon prices faced by domestic and foreign industries, we differentiate between indirect carbon emissions embedded in domestically produced goods, and indirect carbon emissions embedded in imported goods.

---

[13] WIOD reports industry output in monetary units (millions of dollar). Estimating the carbon intensity per monetary unit of output for energy industries requires an approximation of composition of energy industries' output. We follow three steps to estimate the carbon intensity per monetary unit of output for energy industries. First, we source average energy prices by fuel, including coal, gas, oil, diesel and petrol. Second, we compute the carbon intensity per monetary unit of energy industry output by dividing the price per physical unit (e.g. dollar US

In a next step, we transform vector of CO2 emissions per monetary unit of sectors' output ($\boldsymbol{E}_{ind}$) into a vector of CO2 emissions per monetary unit of household expenditure on sectors' output ($\boldsymbol{E}_{indHH}$). This transformation consists of multiple steps, described in section **Error! Reference source not found.**.

To get total carbon emissions per monetary unit of household expenditure, we add indirect to direct emissions $\boldsymbol{E}_{dirHH}$. Direct emissions are released through the consumption of motor and domestic fuels. As HBS data provides expenditure information only, we estimate energy volumes consumed by households by dividing expenditure per fuel by its price[14]. To compute the direct emissions, we multiple the quantity of fuel consumed by its carbon intensity factor, taken from the IPCC 2006 Guidelines for National Greenhouse Gas Inventories (Eggleston et al, 2006). For each household, we add direct and indirect emissions to get final $CO_2$ emissions from household consumption:

$$\boldsymbol{E}_{HH} = \boldsymbol{E}_{dirHH} + \boldsymbol{E}_{indHH} \tag{16}$$

---

per liter) by its carbon content (e.g. 100gCO2 per liter). Third, we use the average fuel mix used by energy industries as weights to produce a single value for the carbon intensity per monetary unit for each energy industry, accounting for the composition of their fuel mix.

[14] A number of studies in the literature calibrate HBS aggregate expenditure to National Account total expenditure. We do not calibrate HBS consumption categories to the HBS in order to avoid changing the composition of households' consumption baskets. In our view, calibration would introduce an additional source of uncertainty around the estimated tax burdens. Calibration would require the assumption that the misreporting and other sources of differences between simulated and official aggregates, such as differences in survey weights used, uprating and imputation techniques (Immervoll and O'Donoghue, 2009), are distributed equally across households. If this were not the case, calibration would introduce bias into the composition of households' baskets. We prefer to let the distributional impact of a tax be driven by the HBS data. We make this choice in view of potential large impacts of calibration on the simulation outcomes (Myck and Najsztub, 2014).

**A.5. Matching HBS to MRIO – an improved approach.**

Cazcarro et al. (2022) follow 4 steps that allow users to transform data from consumption surveys into data from IO models:

1) Align consumption microdata to National Accounts (NA) principles.

    a. Calculate the ratios between Expenditure survey aggregates and Household Final Consumption Expenditure (HFCE) in NA. Use the ratios to adjust consumption in the expenditure survey to HFCE levels.

2) Convert the aligned consumption microdata (in Classification of Individual Consumption by Purpose- COICOP) aligned to Industry classifications (in Classification of products by activity - CPA) using a bridging matrix.

    a. For countries where bridging matrices are unavailable, select a benchmark country as proxy by comparing the structure of expenditure, sociocultural distance, and GDP per capita (Bridging matrices are published in Supplementary materials of Cazcarro et al. (2022).

3) Transform the aligned consumption data (now in NA principles and product-based classifications) from purchaser prices to basic prices.

    a. A tool to do so is published in Supplementary materials of Cazcarro et al. (2022)

4) Adjust the data from the product classification (CPA) to the industry classification.

    a. Use the Supply and Use Tables with the Fixed product sales structure assumption.

The end result will then be a expenditure dataset in Industry terms. Our approach is the reverse of the approach described above.

## A.5. Derivation of committed expenditure and the marginal utility of consumption from the LES

In order to produce compensating variation, a utility function is required. As in the case of Creedy (2001), a Stone-Geary LES direct utility function is utilised:

$$U = \prod_i [\pi_i - \gamma_i]^{\phi_i} \tag{17}$$

where $\gamma_i$ are LES parameters known as committed consumption for each good $i$, $\pi_i$ is the consumption of good i, and $0 \leq \phi_i \leq 1, \sum_i \phi_i = 1$. For convenience ignore the subscripts indicating that different parameters are estimated for different demographic (and income) groups.

Maximising utility subject to the budget constraint $C = \sum_i p_i \gamma_i$, the linear expenditure function for good $i$ is:

$$p_i \pi_i = p_i \gamma_i + \phi_i (C - \sum_i p_j \gamma_i) \tag{18}$$

Differentiating w.r.t. $p_i$ and multiplying by $\frac{p_i}{p_i \pi_i}$, produce the own price elasticity ($\eta_{ii}$) from which the $\gamma_i$ parameters can be derived:

$$\eta_{ii} = \frac{p_i}{p_i \pi_i} - \frac{p_i \phi_i}{p_i \pi_i}(\gamma_i) = \frac{\gamma_i(1-\phi_i)}{\pi_i} - 1 \Rightarrow \gamma_i = \frac{(\eta_{ii}+1)\pi_i}{(1-\phi_i)} \tag{19}$$

Differentiating (**) w.r.t. C and multiplying by $\frac{C}{p_i \pi_i}$, produce the budget elasticity, from which the $\phi_i$ parameters can be derived:

$$\eta_i = \frac{\phi_i C}{p_i \pi_i} \tag{20}$$

Implying:

$$\phi_i = \frac{\eta_i c_i}{C} = \eta_i w_i \tag{21}$$